\documentclass[12pt]{article}
\usepackage{fullpage}
\usepackage{media9}
\addmediapath{./Movies/}
\usepackage{multimedia}
\usepackage{subcaption}
\usepackage[english]{babel}
\usepackage[utf8]{inputenc}
\usepackage{amsmath,amsfonts,amssymb,mathtools,bbm}
\usepackage{algorithm}
\usepackage{booktabs}
\usepackage{caption}
\usepackage{fullpage}
\usepackage{amssymb}
\usepackage{float}
\usepackage{multicol}
\usepackage{dirtytalk}
\usepackage{graphicx}
\usepackage{caption}
\usepackage{adjustbox}
\usepackage{ulem}
\usepackage[margin=1in]{geometry}
\usepackage{graphicx}
\usepackage[colorinlistoftodos]{todonotes}
\usepackage{algorithm}
\usepackage{algpseudocode}
\usepackage{siunitx}
\usepackage{bm}
\usepackage{xcolor}
\usepackage{natbib}
\usepackage{setspace}
\usepackage{color}

\title{Variational Dimension Lifting for Robust Tracking of Nonlinear Stochastic Dynamics}
\author{Yonatan L. Ashenafi}
\date{}

\begin{document}

\maketitle

\begin{abstract}
Nonlinear stochastic motion presents significant challenges for Bayesian particle tracking. To address this challenge, we propose a lifting framework that constructs a higher-dimensional linear stochastic representation of nonlinear state-space models. The resulting surrogates enable the use of computationally efficient linear filtering techniques while retaining a direct connection to the underlying nonlinear dynamics. The paper derives the necessary conditions for such transformations using Ito's lemma and variational calculus, and illustrates the method on a bistable cubic motion model, radial Brownian process model, and a logistic model with multiplicative noise. Simulations confirm that the transformed linear systems, when projected back, accurately reconstruct the nonlinear dynamics and, in distinct regimes of stiffness and singularity, yield tracking accuracy competitive with conventional filters, while avoiding their structural instabilities.
\end{abstract}

\section{Introduction}

Dynamic state estimation in nonlinear stochastic experimental systems remains a central challenge in many areas of science and engineering. In particular, with stochastic differential equations, when nonlinear drift or diffusion terms are present, the distribution of the solution becomes non-Gaussian, making exact tracking with a Kalman filter intractable. Standard approximate filters such as the Extended Kalman Filter (EKF) and Unscented Kalman Filter (UKF) perform approximations either by linearizing the model (EKF) or by propagating a small set of points that encode only the mean and covariance of the state distribution (UKF), and thus may perform poorly when nonlinearities are strong or the system exhibits transitions between different dynamical regimes \citep{julier1997new, simon2006optimal}.

A different strategy is to transform the representation of the dynamics. Operator-theoretic and lifting-based frameworks, such as those based on Koopman operators or Carleman linearization, provide a rigorous foundation for global linearization and dimensional reduction \citep{brunton2016koopman, Mauroy2019, Klus2020, gutierrez2021reduced}. While these methods extend naturally to stochastic systems \citep{Crnjaric2017, Nueske2023}, they face significant practical hurdles. Because these frameworks are theoretically infinite-dimensional, finite-dimensional truncations often introduce closure errors and necessitate high-dimensional approximations that might be computationally expensive. Furthermore, they generally do not guarantee the invertibility of the lifting map or strict consistency with Itô calculus. In parallel, Gaussian variational surrogates approximate time-marginal densities \citep{Archambeau2007}, but because they prioritize marginals over a direct representation of the underlying dynamics, they can complicate sequential filtering. 

This work develops a framework that constructs an It\^{o}-consistent representation of a nonlinear stochastic system in which the dynamics become linear in a finite-dimensional space. The central idea is to determine both the lifting map and the associated linear dynamics so that the lifted evolution satisfies the It\^{o} differential rules, while preserving the ability to reconstruct the original state. We formulate this as a variational problem that enforces It\^{o}-consistency while weighting approximation error according to the stationary distribution of the original process, emphasizing fidelity where the dynamics would concentrate.

This paper is organized into three main parts. First, we introduce a variational formulation to construct a finite-dimensional linear surrogate for nonlinear SDEs. By minimizing an objective functional weighted by the system's stationary density, we derive a coordinate transformation that is globally consistent with It\^{o} calculus. Second, we apply this framework to three distinct stochastic processes—cubic bistable motion, radial Bessel diffusion, and Wright-Fisher logistic evolution—and analyze how the spectral properties of the optimized linear operator impact the long-term stability of the surrogate. Third, we leverage these lifted models for Bayesian particle tracking, demonstrating how that Kalman filtering in the lifted coordinates compares with EKF, UKF, particle filtering (Sequential Monte Carlo).

This approach provides a practical middle ground between Gaussian approximations, operator-theoretic infinite-dimensional embeddings, and purely approximation-based filtering, yielding interpretable linear surrogates of nonlinear stochastic processes with identifiable regimes of validity.

\section{Background}
Continuous-time state space models (SSMs) are widely used to describe a given system's dynamics. Stochastic Differential Equations (SDEs) are the favored choice for constructing these models. The SSMs are given by a motion model and an observation model. For simplicity, we will have non-dimensional equations. The motion model can be given by
 \begin{equation} \label{Eqn_general_SDE}
 \begin{aligned}
 d \mathbf{x}= \mathbf{f}( \mathbf{x})dt+ \mathbf{G}( \mathbf{x}) \mathbf{dW}_t
  \end{aligned} 
  \end{equation}

where 
\(\mathbf{x}(t) \in \mathbb{R}^n\) is the latent/hidden state, 
\(\mathbf{f} : \mathbb{R}^n \to \mathbb{R}^n\) is the drift function describing deterministic dynamics,  
\(\mathbf{G} : \mathbb{R}^n \to \mathbb{R}^{n \times m}\) is the diffusion matrix describing how noise enters the system, and  
\(\mathbf{W}_t \in \mathbb{R}^m\) is an \(m\)-dimensional standard Wiener process.  

The observation model relates the latent states to measurable quantities:
\begin{equation} \label{Eqn_general_obs}
\mathbf{y}(t) = \mathbf{h}(\mathbf{x}(t)) + \mathbf{v}(t),
\end{equation}
where \(\mathbf{y}(t) \in \mathbb{R}^p\) is the observation vector,  
\(\mathbf{h} : \mathbb{R}^n \to \mathbb{R}^p\) is the observation function, and  
\(\mathbf{v}(t)\) represents observation noise, often modeled as a Gaussian process.  

These SSM models serve as the foundation for state prediction via methods like the Kalman filter, Extended Kalman filter (EKF), Unscented Kalman filter (UKF), and particle filters (Sequential Monte Carlo). The classical Kalman filter provides exact solutions when both the motion and observation models are linear and Gaussian whereas the EKF, UKF, and other advanced methods try to give good estimates by approximation approaches. They approximate the true Bayesian update of the actual SSM model. The framework in this paper offers an alternative perspective. Instead of directly approximating the nonlinear dynamics, it transforms the original state space into a finite higher-dimensional feature space where the dynamics can be represented as linear. 

\section{Methods} 
We want to transform particles' motion and/or observation models to get a linear SSM. We start with the general system of SDEs on $\mathbb{R}^N$ for the latent states $\mathbf{x}$.
 \begin{equation} \label{Eqn_general_SDE}
 \begin{aligned}
 d \mathbf{x}= \mathbf{f}( \mathbf{x})dt+ \mathbf{G}\mathbf{dW}_x
  \end{aligned} 
  \end{equation} 
  Throughout this paper, we treat the system dynamics in dimensionless units. All state variables $x$ and time parameters $t$ are normalized such that the characteristic length and time scales of the motion are of order unity. Consequently, physical units are omitted.
  
  We now define the transformation $T: \mathbf{x}\rightarrow  \mathbf{U}(\mathbf{x})=(U_1( \mathbf{x}),U_2( \mathbf{x}),...,U_M( \mathbf{x}))$ that leads to a linear SSM given by:
 \begin{equation} \label{Eqn_general_SDE_transformed}
 \begin{aligned}
 d \mathbf{U}= \mathbf{AU}dt+ \mathbf{BUdW}^e_t
  \end{aligned} 
  \end{equation}
Where $A$ and $B$ are matrices in $\mathbb{R}^{MxM}$. $\mathbf{dW^e_t}=[\mathbf{dW_t},\mathbf{0}]$. We also assert that the transformation T is invertible and therefore that $U_i$ is a bijective function for each $i$.
Next we use Ito's lemma to write down the equations that govern the transformation \citep{kloeden_platen_1992}. 
 \begin{equation} \label{Eqn_ItoCond_1}
 \begin{aligned}
  (\nabla U_i)^T(\mathbf{f})+\frac{1}{2}Tr[\mathbf{G}^TH(U_i)\mathbf{G}]= \sum\limits_{j=1}^{M}A_{i,j}U_j
  \end{aligned} 
  \end{equation}
\begin{equation} \label{Eqn_ItoCond_2}
 \begin{aligned}
  (\nabla U_i)^T(\mathbf{G})= \mathbf{e}_i^T\mathbf{B}
  \end{aligned} 
  \end{equation}
Where $H(U_i)$ is the Hessian of $U_i$. (\ref{Eqn_ItoCond_1}) represents a system of second order linear PDEs and (\ref{Eqn_ItoCond_2}) is a system of first order PDEs that is an element of an array of such systems. We assume that G is a constant orthonormal matrix. This assumption corresponds to the regime of additive isotropic noise, which is standard in overdamped Langevin models of molecular dynamics and cellular migration where thermal fluctuations are homogeneous. The Ito condition will then be

 \begin{equation} \label{Eqn_ItoCond_case1a}
 \begin{aligned}
  \mathbf{J}\mathbf{f}(\mathbf{x})+\frac{1}{2}\begin{pmatrix}
\operatorname{Tr}\big[G^T H(U_1) G\big] \\
\vdots \\
\operatorname{Tr}\big[G^T H(U_M) G\big]
\end{pmatrix}= \mathbf{A U}
  \end{aligned} 
  \end{equation}
   \begin{equation} 
 \begin{aligned}
 \mathbf{J}\mathbf{G}\mathbf{dW_x}= (\mathbf{BdW^e_x})^T
  \end{aligned} 
  \end{equation}
Where $J$ is the Jacobian of $\mathbf{U}$.

Define the infinitesimal generator acting on a scalar \(u\) by
\[
\mathcal{L}u(x) := f(x)u'(x) + \tfrac12 G(x)^2 u''(x).
\]
Let
\[
R(x;U,A) := \mathcal{L}U(x) - A\,U(x)\in\mathbb{R}^M,
\qquad
S(x;U,B) := G(x)U'(x) - B\,U(x)\in\mathbb{R}^{M_W}. 
\]
and define the objective functional
\begin{equation}\label{eq:J-continuous}
\mathcal{J}[U,A,B] \;=\; \int_{\Omega} \Big( \|R(x;U,A)\|_2^2 + \|S(x;U,B)\|_2^2 \Big)\, \rho(x)\,dx.
\end{equation}

Here we choose to be the stationary distribution of the original SDE. While in this theoretical study we use the analytical  to rigorously benchmark performance, in practical experimental settings where is unknown, it can be approximated via a short 'burn-in' simulation or estimated iteratively (e.g., recursive density estimation). This weighting strategy is theoretically grounded in the ergodic property of the stochastic process \citep{pavliotis2014stochastic}. Under the assumption of ergodicity, the Birkhoff Ergodic Theorem establishes the equivalence between the ensemble average over the state space and the long-term time average of a single trajectory:
\[
    \lim_{T \to \infty} \frac{1}{T} \int_{0}^{T} \mathcal{E}(X_t) \, dt = \int_{\Omega} \mathcal{E}(x) \rho(x) \, dx,
\]
where $\mathcal{E}(x)$ represents the local approximation error (the integrand of $\mathcal{J}$). Consequently, minimizing the objective functional weighted by $\rho(x)$ is equivalent to minimizing the time-averaged residual error accumulated by the particle along its physical trajectory. Our goal is to minimize \(\mathcal{J}\) jointly over \(U,A,B\). This follows the broader idea of enforcing operator constraints through residual minimization, as used in physics-informed neural networks \citep{raissi2019physics}. The general Euler-Lagrange equations governing the optimal lifting map $U$, derived from the first variation of $\mathcal{J}$, are presented in the Appendix. These equations form the theoretical basis for a future alternating optimization scheme in which $U$ is solved numerically given $A$ and $B$, and $(A, B)$ are re-optimized given $U$.

One interesting modification is to apply the stationary distribution of the lifted system instead of or along with the stationary distribution of the original SDE in the objective. The stationary distribution of the original system prioritizes accuracy at frequently visited regions of the original system while the stationary distribution of the lifted system prioritizes accuracy at frequently visited regions of the lifted system. Here we pick the former approach and leave the later approach for future work.

Before we proceed with the optimization we notice that the objective being minimized is only relevant when the lifted dynamics stays within the bounds of integration. That is why have the other objective of controlling the spectrum of A. In particular we want the eigenvalue with the largest real part to have a real part that is negative or as small as possible. 

We do not enforce stability of $A$ through a parametrization that guarantees it a priori, because excursions of the lifted process outside the integration domain at large times are not problematic for our intended use case. Ultimately, our goal is to track trajectories of the original/latent system over a finite observational window. Therefore, the degree to which we penalize the spectral radius of A can be tuned depending on the time scales relevant to the specific tracking problem under consideration.

The objective now becomes 
\begin{equation}
\begin{aligned}
    & \min_{U, A, B} \quad \mathcal{J}[U,A,B] + \mu_{\mathrm{stab}}\,\phi\big(\alpha(A)\big) \\
    & \text{subject to} \quad U_1(x) = x,
\end{aligned}
\label{eq:constrained_optimization}
\end{equation}
where $\alpha(A):=\max_i \Re(\lambda_i(A))$, $\phi(z):=(\max\{0,z\})^2$, $\mu_{\mathrm{stab}}$ balances fidelity to the Itô constraints against stability of the lifted linear system. The anchor constraint $U_1(x)=x$ guarantees that the latent state is directly embedded within the lifted coordinates, providing straightforward reconstruction.

To interpret the magnitude of the residual $\mathcal{J}$ across systems with different physical scales, we introduce a normalized metric of model fidelity. We define the lifting coefficient of determination, $R^2_{\text{lift}}$, by comparing the optimized residual against a baseline model that assumes static dynamics ($A=0, B=0$):
\begin{equation}
    R^2_{\text{lift}} = 1 - \frac{\mathcal{J}(U, A, B)}{\mathcal{J}_{\text{null}}}, \quad \text{where } \mathcal{J}_{\text{null}} = \int_{\Omega} \left( \|\mathcal{L}U\|^2 + \|\nabla U (\nabla U)^T\|^2 \right) \rho(x) \, dx.
    \label{eq:r_squared}
\end{equation}
An $R^2_{\text{lift}}$ approaching 1 indicates that the linear surrogate successfully captures the dominant drift and diffusion characteristics of the original process, while a value near 0 implies the linear model offers no predictive improvement over a static estimate.

The constant-$G$ simplification in Eqs.~(7)--(8) applies to Sections~3.1 and~3.2. For Sections~3.3 where we have state-dependent diffusion $g(x)$, the framework is extended to multiplicative noise by retaining the full state-dependent generator. This constitutes an acknowledged approximation: the theoretical It\^o conditions~(7)--(8) are replaced by their state-dependent analogues, and the resulting lifted system is linear with fixed $B$ despite the original process having state-dependent noise.

We demonstrate the framework on three canonical one-dimensional stochastic processes that together span additive noise with bistability, singular drift, and multiplicative noise.

For all three test cases, $A_0 = -0.5\,I$ and $B_0 = I$. The exponents are initialized as follows:
$\alpha_0 = [0.8, 1.0, 1.2]$ for the Wright--Fisher case;
$\alpha_0 = [0.0, 0.1, -0.1]$ for the Bessel case; and
$\alpha_0 = [0.05, -0.05, 0.10]$ for the cubic bistable case. For all three we set $\mu_{stab}=1$. Sensitivity to initialization was verified by running 50 trials with exponents drawn uniformly from $[-2, 2]$ and Gaussian perturbations of magnitude $0.1$ added to $A_0$ and $B_0$. For the Wright--Fisher and Bessel cases, all 50 trials converged to the same loss value within $\pm 0.1\%$. For the cubic bistable case, 44 of 50 trials converged to the near-zero local minimum at $\mathcal{J} = 2.643$, while 6 trials found the better basin at $\mathcal{J} \approx 0.086$.

\subsection{Cubic Bistable Process}
A fundamental example of a nonlinear stochastic system with bistability is the cubic drift process \citep{kramers1940Brownian}. The SDE is given by
\begin{equation}\label{eq:conreteSDE}
dx=-x(x-1)(x+1)dt+\sigma dW
\end{equation}
Our goal is to minimize \(\mathcal{J}\) jointly over \(\alpha, \beta, \gamma,A,B\). To do this we find the stationary distribution $\rho$ first. It is given by
\[
\rho(x) = \frac{1}{Z} \exp\!\left(\frac{2}{\sigma^2}\left(-\frac{x^4}{4} + \frac{x^2}{2}\right)\right),
\]
where $Z$ is the normalizing constant. The distribution is bimodal with probability mass concentrated near $x = \pm 1$, as expected for a bistable system.

Motivated by the fact that exponential functions are eigenfunctions of constant-coefficient linear operators, we consider the trial functions $\mathbf{U}(x) = (x, e^{\alpha x}, e^{\beta x}, e^{\gamma x})^\top$. While simple, this basis is connected to the Moment Generating Function (MGF), $M_x(s) = \mathbb{E}[e^{sx}]$. By predicting the evolution of $\mathbb{E}[e^{\alpha_i x}]$ for optimized exponents $\alpha_i$, our linear surrogate effectively tracks the time-evolution of the original variable's probability density through its characteristic features. The ``anchor'' $U_1(x) = x$ ensures that the first moment (the physical state) is always explicitly represented.

For the optimization problem we set $\sigma = 2$ and evaluate 
the integral over $x \in [-10.0, 10.0]$ using steps of size 
$dx = 0.005$. A multiple initialization study found a solution 
with $\mathcal{J} = 0.086$ and $R^2_{\text{lift}} = 0.991$, 
but the corresponding $(A, B)$ matrices have entries of order 
$10^4$, causing numerical instability when the continuous-time 
lifted system is subsequently discretized for filtering. Another 
solution, while achieving a higher $\mathcal{J} = 2.643$ and 
$R^2_{\text{lift}} = 0.73$, has well-conditioned $(A, B)$ 
matrices with entries of order one and is therefore retained 
for all tracking experiments. This finding illustrates the optimal matrices $(A, B)$ must also yield a numerically stable discretization. This imposes an implicit constraint on the allowable forms of $A$ and $B$.

With the above selected solution with $J = 2.643$ we obtain:

\[
\begin{aligned}
(\alpha,\beta,\gamma)
&\approx \big(-9.53 \times 10^{-8},\; -1.57 \times 10^{-8},\; -8.59 \times 10^{-8}\big), \\[6pt]
A &=
\begin{pmatrix}
-1.550 & 0.625 & 0.630 & -1.255 \\
1.36 \times 10^{-7} & -2.478 & -0.138 & 2.616 \\
3.97 \times 10^{-8} & -2.267 & -0.910 & 3.177 \\
3.32 \times 10^{-8} & -1.999 & 0.697 & 1.302
\end{pmatrix}, \\[8pt]
B &=
\begin{pmatrix}
1.78\times10^{-7} & 1.28 & 8.26\times10^{-1} & -1.11\times10^{-1} \\
-1.79\times10^{-7} & 1.20 & -2.26\times10^{-1} & -9.75\times10^{-1} \\
-3.63\times10^{-7} & 7.69\times10^{-1} & 8.81\times10^{-1} & -1.65 \\
-2.14\times10^{-7} & 1.58 & -9.47\times10^{-2} & -1.49
\end{pmatrix}.
\end{aligned}
\]

\begin{figure}[H]
    \centering
    \includegraphics[scale=0.5]{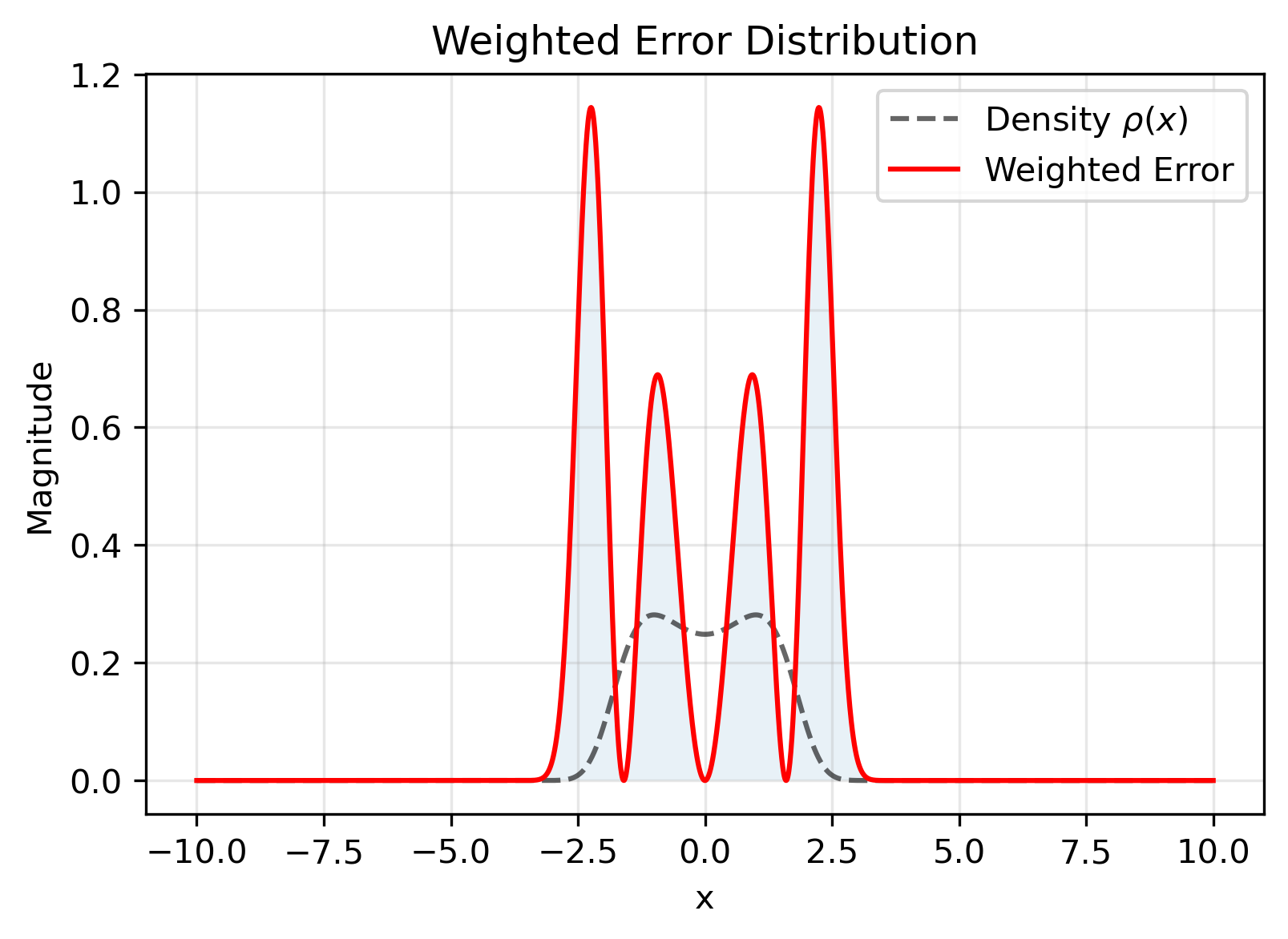}
    \includegraphics[height=5cm]{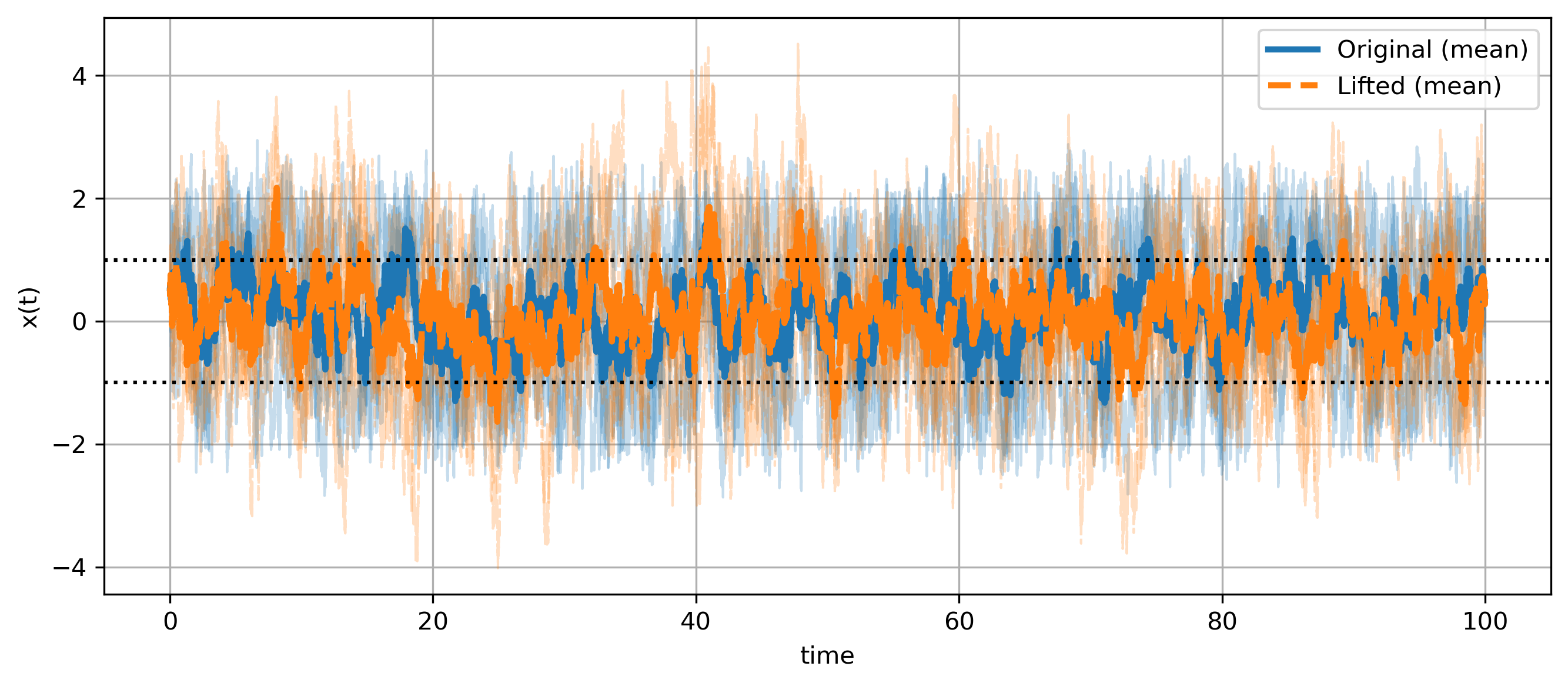}
    \caption{
    Optimization diagnostics for the cubic bistable diffusion ($M=4$). Top: The stationary density $\rho(x)$ (black dashed line) and the spatial distribution of the weighted residual error (red solid line). Bottom: Simulated dynamics of the original versus lifted systems (with $dt=0.001$, $x(0)=0.5$, $\mu_{\text{stab}}=1$). We used 5 trajectories of each for the comparison.
    The lifted dynamics closely reproduces the bistable behavior of the original SDE.
    }
    \label{fig:lagrangian_compare}
\end{figure}

The rightmost eigenvalue of A is found to be $\lambda \approx -8.69\times 10^{-7}$. 
The integrand in \(\mathcal{J}\) and the dynamics comparison look as shown in figure \ref{fig:lagrangian_compare}.  

As a sensitivity analysis on the regularization parameter $\mu_{stab}$, we performed an ablation study varying $\mu_{stab}$ over two orders of magnitude. The results are given in table \ref{tab:mu_ablation}.
\begin{table}[H]
\centering
\begin{tabular}{lccc}
$\mu_{\mathrm{stab}}$ & $\mathcal{J}$ & $R^2_{\mathrm{lift}}$ & $\max \Re(\mathrm{eig}(A))$ \\
\hline
0.0  & 2.6431 & 0.7287 & $8.37\times 10^{-7}$ \\
0.1  & 2.6431 & 0.7287 & $-3.40\times 10^{-6}$ \\
1.0  & 2.6431 & 0.7287 & $-8.68\times 10^{-7}$ \\
10.0 & 2.6431 & 0.7287 & $-5.09\times 10^{-5}$ \\
\end{tabular}
\caption{Effect of $\mu_{\mathrm{stab}}$ on the objective value $\mathcal{J}$, lifted $R^2$, and spectral stability.}
\label{tab:mu_ablation}
\end{table}

\subsection{Radial (Bessel) Process}
Another important example of a nonlinear stochastic system is the radial (Bessel-type) process \citep{borodin2002handbook}. Let 
\[
\mathbf{B}(t) = \big( B(t)^{(1)}, B(t)^{(2)}, \dots, B(t)^{(n)} \big)
\]
be standard Brownian motion in $\mathbb{R}^n$.  
The radial part is defined as
\[
R(t) = \|\mathbf{B}(t)\| = \sqrt{ (B(t)^{(1)})^2 + \cdots + (B(t)^{(n)})^2 }.
\]
Applying It\^o's formula, the dynamics of $R(t)$ satisfy the stochastic differential equation
\begin{equation}
dR = \frac{n-1}{2R} \, dt + \sigma dW_t,
\label{eq:bessel_process}
\end{equation}
where $W_t$ is a standard one-dimensional Brownian motion. We impose reflective boundary conditions at $R=R_{max}$. From the Fokker Planck equation we find the stationary density to be 
\[
\rho(R) = \frac{n R^{\,n-1}}{R_{\max}^{\,n}}, 
\qquad 0 \le R \le R_{\max}.
\]

We set $R_{\max}=5$ and $n=3$. Note that, in this case, the reflecting boundary condition does not appear explicitly in the linear SDE parameters. Its effect enters solely through (i) the stationary density $\rho$ used to weight the objective, and (ii) the finite interval over which the objective integral is evaluated.

\begin{figure}[H]
    \centering
    \includegraphics[height=3.8cm]{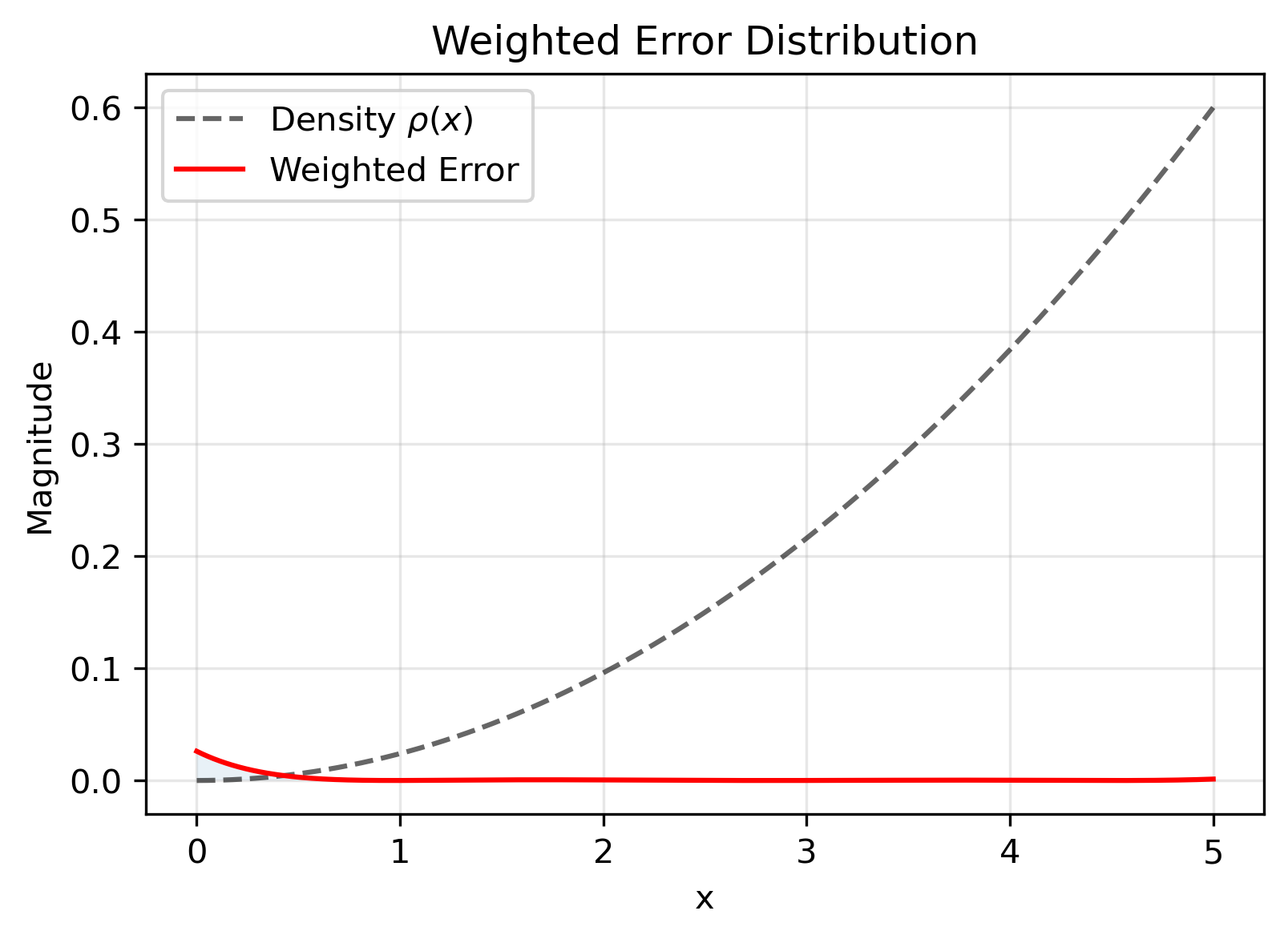}\\
    \includegraphics[height=3.5cm]{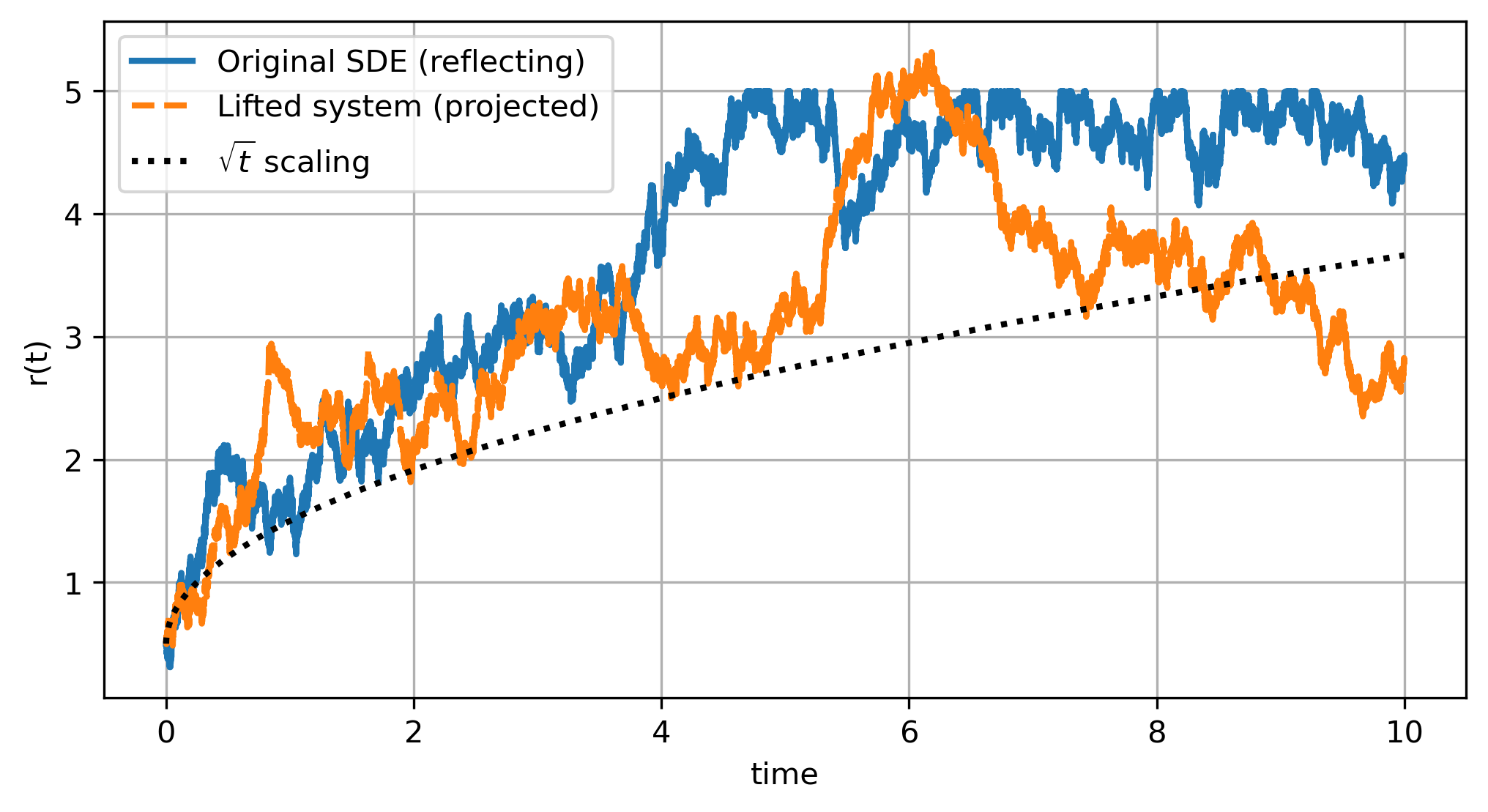}
    \includegraphics[height=3.5cm]{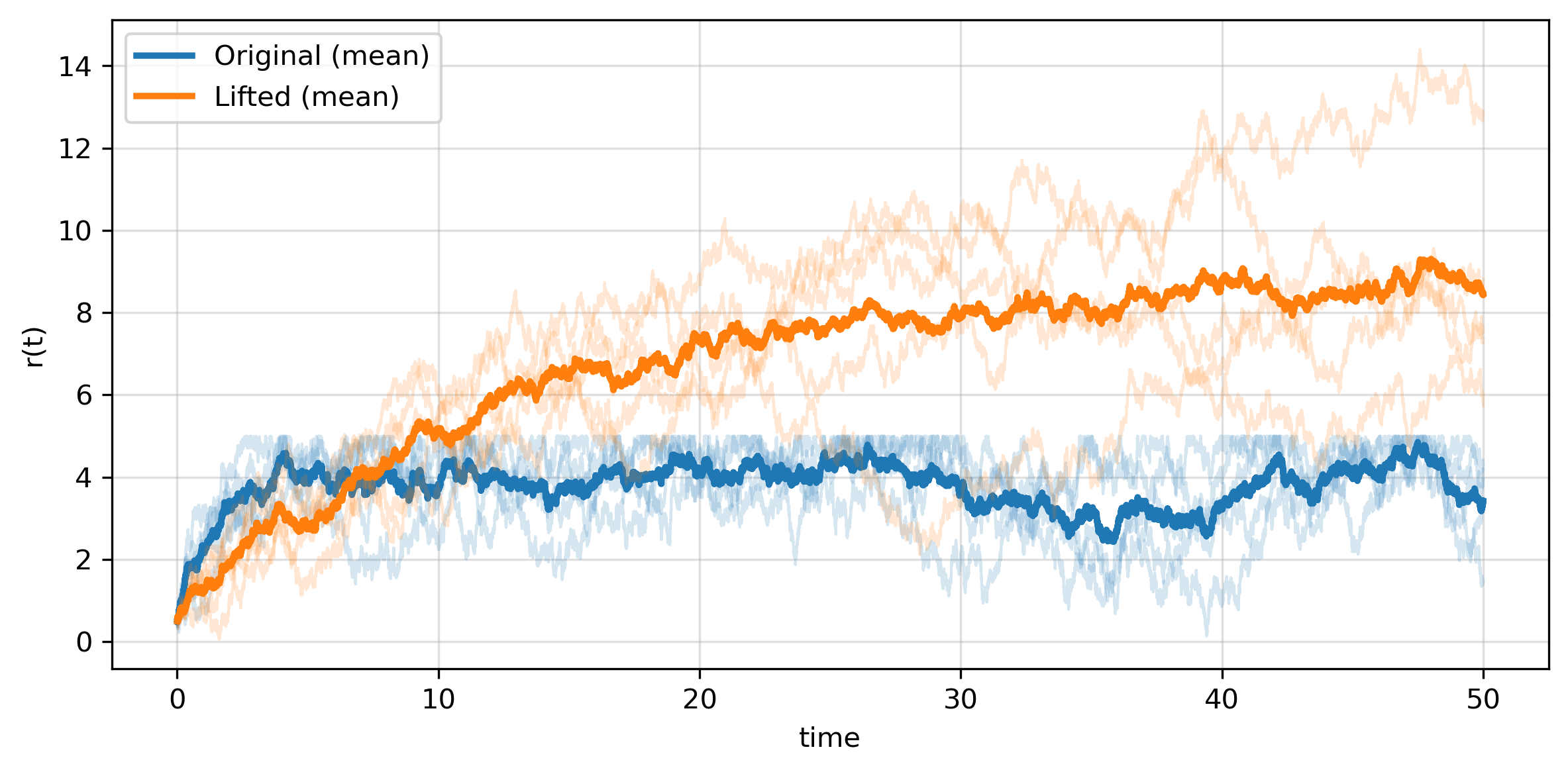}
    \caption{
   Top: The stationary density $\rho(x)$ and the spatial distribution of the weighted residual error. Bottom: The simulated dynamics of the original versus lifted systems  with $dt=0.001$, $x(0)=0.5$. We used 5 trajectories of each for the comparison. The original dynamics grows at an order of $\sqrt{t}$, as expected. 
The lifted dynamics reproduces the growth behavior of the original SDE's solution over the time frame considered. 
    }
    \label{fig:lagrangian_compare_radial}
\end{figure}

Here too we use three exponential basis functions and use the \texttt{BFGS} algorithm to solve for \(\alpha, \beta, \gamma,A,B\). We set $\sigma=1$ and $\mu_{\text{stab}}=1$. For the integral we evaluate over $r \in[10^{-8}, R_{\max}]$ using steps of size $dr=0.0005$. We get the rightmost eigenvalue of A having a real part $\lambda \approx  -0.0075$. We also obtain: 
\[
(\alpha,\beta,\gamma) =
\begin{pmatrix}
0.0726, \;
-0.284, \;
0.0119
\end{pmatrix}.
\]
\[
A =
\begin{pmatrix}
0.252 & 1.39 & 4.11 & -3.78 \\
-0.0180 & 0.283 & 0.125 & -0.307 \\
-0.193 & 0.765 & -1.29 & 0.154 \\
-0.0526 & 0.648 & 0.00853 & -0.627
\end{pmatrix},
\hspace{1 em}
B =
\begin{pmatrix}
-0.108 & 0.522 & -0.428 & 0.833 \\
-0.0220 & 0.156 & -0.123 & 0.0165 \\
0.216 & -2.02 & 0.142 & 1.62 \\
0.0702 & -0.667 & 0.131 & 0.555
\end{pmatrix}
\]

\(R^2_{\text{lift}}\) reaches a value of 0.99. The $\rho$-weighted error is shown in in figure \ref{fig:lagrangian_compare_radial}.  


\subsection{Logistic-type diffusion with mutation} \label{sec:Logistic}

As a multiplicative-noise example on a bounded domain, we consider a
Wright--Fisher-type diffusion. In this process the state represents a fraction variable confined to
the unit interval starting at zero. Examples include the proportion of active/diffusive molecular motors and the occupancy of a biochemical site. It is widely used in its generator (Kolmogorov equation) form \citep{ewens2004mathematical, wakeley2009coalescent}. The SDE is given by
\begin{equation}
  dX_t = \kappa\big(\theta_1(1-X_t)-\theta_0 X_t\big)\,dt
         + \sqrt{2\kappa\,X_t(1-X_t)}\,dW_t,
  \label{eq:wf_sde}
\end{equation}
with parameters $\kappa>0$ and $\theta_0,\theta_1>0$. The drift term
\[
  \mu(x) = \kappa\big(\theta_1(1-x)-\theta_0 x\big)
\]
combines growth and saturation with mutation-like
fluxes from the boundaries, while the diffusion coefficient
\[
  g^2(x) = 2\kappa x(1-x)
\]
is multiplicative and vanishes at $x=0$ and $x=1$.

A stationary density $\rho$ with zero probability flux satisfies
\[
  \mu(x)\rho(x) - \frac{1}{2}\frac{d}{dx}\big(g^2(x)\rho(x)\big) = 0.
\]
Using the standard one-dimensional formula, we obtain
\[
  \rho(x) 
  = C\,x^{\theta_1-1}(1-x)^{\theta_0-1}, \qquad x\in(0,1).
\]
Normalizing yields a Beta distribution
\begin{equation}
  \rho(x) = \frac{1}{B(\theta_1,\theta_0)}
  x^{\theta_1-1}(1-x)^{\theta_0-1}, \qquad x\in(0,1),
  \label{eq:wf_stationary}
\end{equation}
with $\theta_0,\theta_1>0$. Thus the logistic-type diffusion with mutation
provides a bounded, multiplicative-noise test case with an
explicit stationary density.

Here too we use three exponential basis functions and use the \texttt{BFGS} algorithm to solve for \(\alpha, \beta, \gamma,A,B\). We set $\kappa=2$ and set the domain of integration at $x \in [0,1]$ with $dx=0.0005$. Unlike the cubic bistable case, the near-zero exponents for the Wright--Fisher process represent the true global minimum of $\mathcal{J}$ rather than a local attractor: all 50 random initializations converge to $\mathcal{J} = 0.0088$, $R^2_{\text{lift}} = 0.9997$.  However, at $|\alpha_i| \sim 10^{-5}$ the basis functions 
$e^{\alpha_1 x}, e^{\alpha_2 x}, e^{\alpha_3 x}$ are nearly 
identical to each other on $[0,1]$, 
so the optimizer cannot distinguish between many different 
$(A, B)$ pairs that produce the same $\mathcal{J}$. The reported 
$(A, B)$ matrices are therefore not unique. Finding a basis 
that achieves low $\mathcal{J}$ while keeping the basis functions 
genuinely distinct is an important direction for future work.

We get the rightmost eigenvalue of A has a real part $\lambda \approx -1.25*10^{-7}$. The remaining results are given here: 
\[
(\alpha,\beta,\gamma) =
\begin{pmatrix}
-1.18 \times 10^{-5}, \;
-1.02 \times 10^{-5}, \;
-8.03 \times 10^{-6}
\end{pmatrix}
\]
\[
A =
\begin{pmatrix}
-14.0 & 4.37 & 3.35 & 2.29 \\
1.65 \times 10^{-4} & -0.205 & 0.231 & -0.0257 \\
1.44 \times 10^{-4} & 0.443 & -0.243 & -0.200 \\
1.16 \times 10^{-4} & 0.701 & 0.329 & -1.03
\end{pmatrix}, 
B =
\begin{pmatrix}
-0.870 & 0.546 & 0.561 & 0.342 \\
1.14 \times 10^{-5} & 0.475 & -0.380 & -0.0955 \\
8.65 \times 10^{-6} & -0.517 & 0.638 & -0.120 \\
3.82 \times 10^{-6} & -0.555 & -0.387 & 0.942
\end{pmatrix}
\]

We note that no finite-dimensional linear lift with constant $B$ can exactly satisfy the Itô conditions for the WF diffusion 
$g(x) = \sqrt{2 \kappa x (1-x)}$, because exact satisfaction of the diffusion condition would require all basis functions $U_i$ to share the same derivative structure 
$U_i'(x) \propto 1/g(x)$, precluding a linearly independent lifting basis. The framework instead finds the best $L^2$ approximation under the stationary measure, 
and the resulting $R^2_\text{lift} = 0.998$ confirms that this approximation is tight for the parameter values used.

\begin{figure}[H]
    \centering
    \includegraphics[scale=0.5]{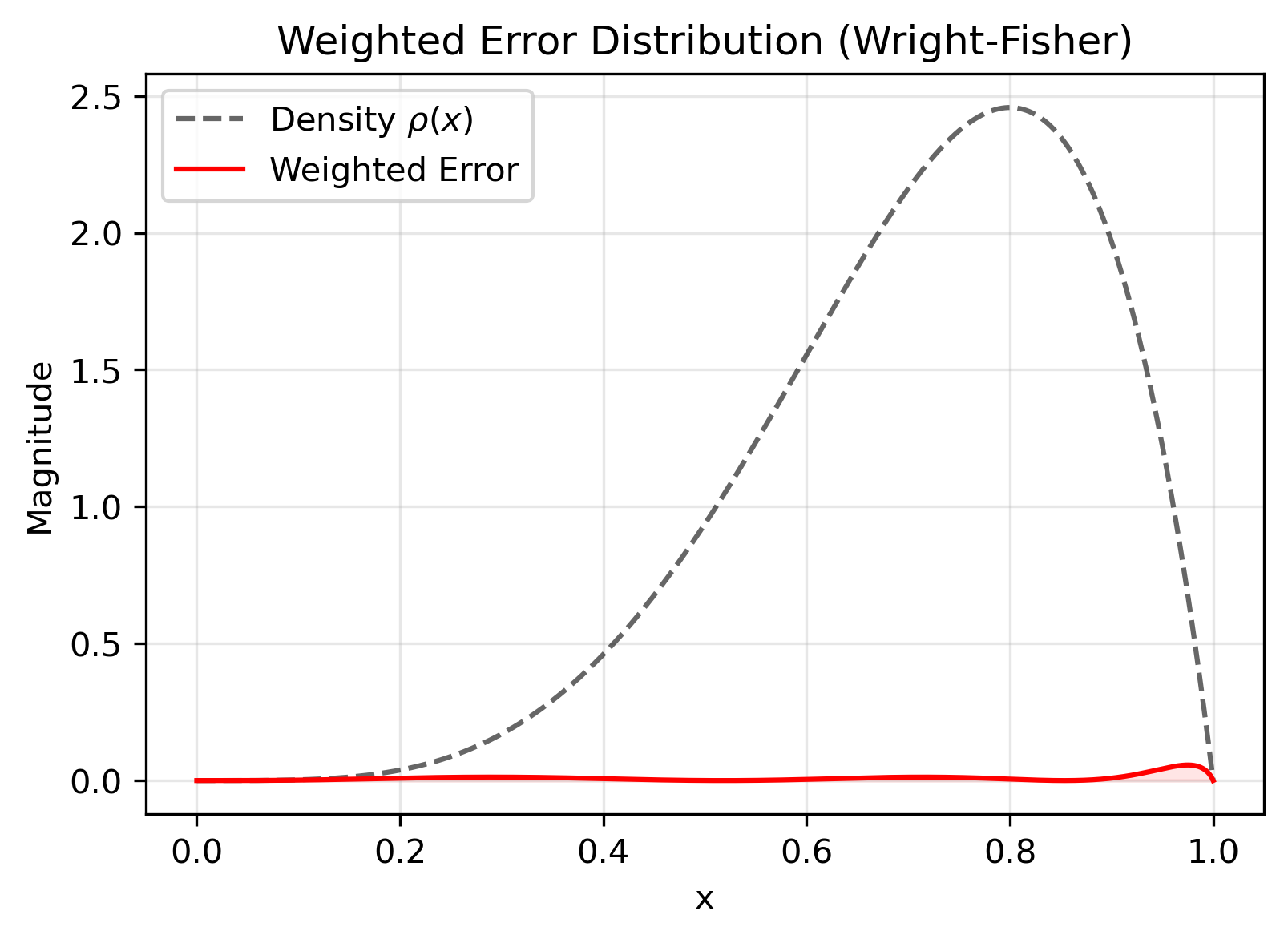}\\
    \includegraphics[scale=0.5]{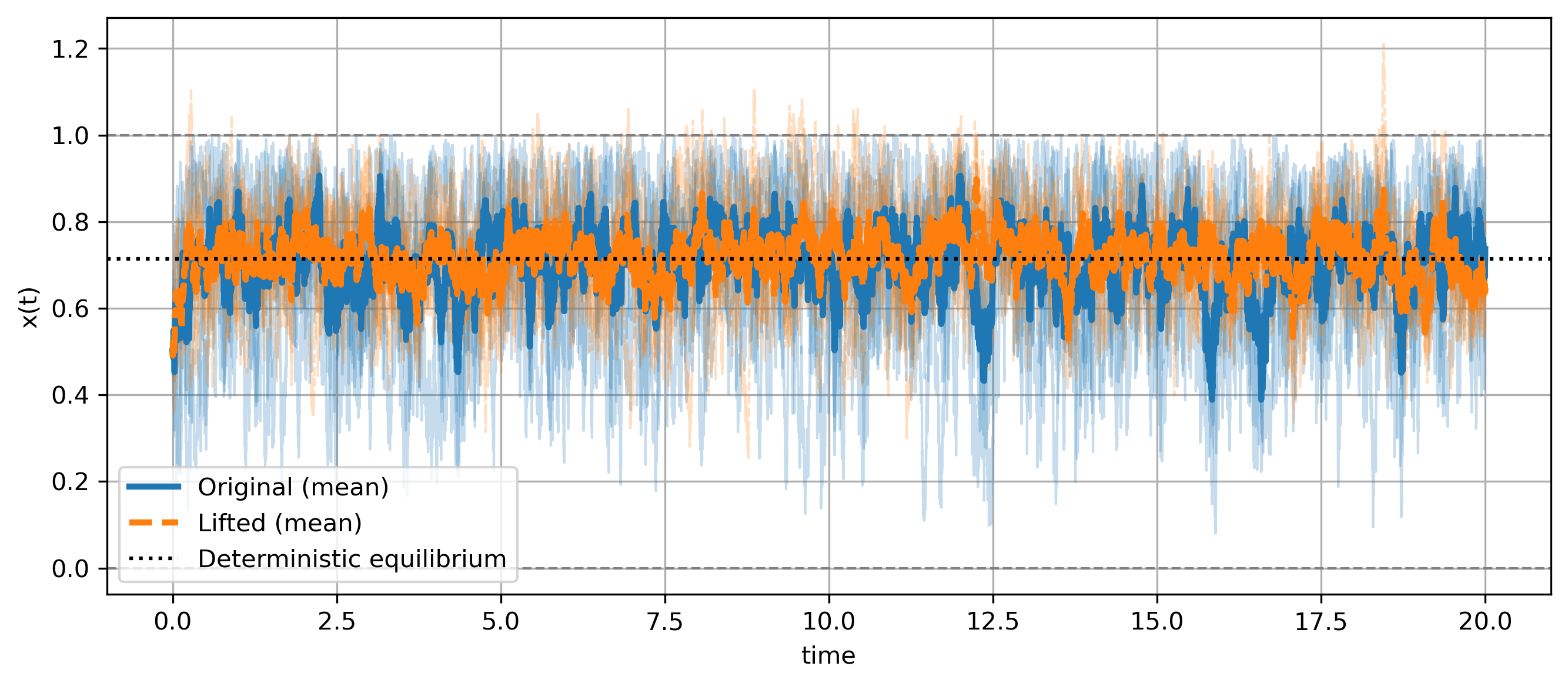}
    \caption{Top: The stationary density $\rho(x)$ and the spatial distribution of the weighted residual error. Bottom: Simulated dynamics of the original versus lifted systems ($dt = 0.001$, $x(0) = 0.5$). The lifted dynamics closely reproduces the stationary distribution and mean trajectory of the original SDE. In particular, the lifted dynamics stays near the deterministic stable equilibrium of the original system over the time frame displayed.}
    \label{fig:lagrangian_compare_logistic}
\end{figure}

The $\rho$-weighted error is shown in in figure \ref{fig:lagrangian_compare_logistic}.   

\section{Application to Particle Tracking}

Particle tracking experiments typically yield discretely sampled trajectories of diffusing or actively transported particles, with measurement noise arising from various sources. Classical analyses treat such trajectories as noisy observations of underlying stochastic differential equations \citep{berglund2010statistics}. Our simulation and filtering
experiments adopt the same viewpoint: an underlying continuous-time SDE governs the particle’s motion, which is discretely observed with noise. 

To apply standard linear-Gaussian inference, we must discretize the continuous lifted system $dU = AU dt + BU dW$ derived in previous sections. We formulate a discrete-time SSM of the form:
\begin{equation}
    \begin{aligned}
        U_{k+1} &= F U_k + w_k, & w_k &\sim \mathcal{N}(0, Q) \\
        y_k &= C U_k + v_k, & v_k &\sim \mathcal{N}(0, R)
    \end{aligned}
\end{equation}
where $U_k$ is the lifted state at time $t_k = k\Delta$ and $y_k$ is the scalar observation. 

Although the motivation arises from particle-tracking studies, the methodology developed here applies equally to any setting where a nonlinear SDE is observed at discrete times.
 
\subsection{Cubic Bistable Process}
The SDE is given in equation \ref{eq:conreteSDE}. We generate trajectories with a small step $dt$ (e.g., $dt=10^{-3}$), sample at interval $\Delta$ (e.g., $\Delta=0.1$), and add Gaussian observation noise $\sigma_y=0.25$:
\[
  x_{t+dt} = x_t - x_t(x_t-1)(x_t+1)dt + \sigma_x \sqrt{dt}\, \xi_t,\;\; \xi_t\sim\mathcal N(0,1),
  \quad 
  y_k = x_{t_k} + \varepsilon_k, \;\; \varepsilon_k \sim \mathcal N(0, \sigma_y^2).
\]

We initialize $x_0$ from the stationary density of the cubic process.
To rigorously evaluate the robustness of the method, we introduce a discrepancy between the data-generating process and the filter's prior assumptions. Specifically, the synthetic trajectories are generated with a diffusion coefficient of $\sigma_{\text{true}}=1$, while the linear surrogate model is optimized using a baseline assumption of $\sigma_{\text{model}}=2$. This mismatch serves as a stress test, simulating a realistic experimental scenario where the underlying physical parameters are imperfectly known. 

We apply the Kalman filter to $\{U_k\}$ using $(F,Q,C,R)$. We initialize $U_0$ by mapping $x_0$ to $U(x_0)$ and assigning a conservative initial covariance $\Sigma_{U,0}$ (diagonal with modest variance terms). We use sparse observations ($\Delta = 0.4$). The
process exhibits long dwell-times in either potential well, punctuated by rare transitions.
 
We compare our simple lifting method with UKF, EKF, Sequential Monte Carlo, and regular Kalman filtering.
For benchmarking, we configured the EKF using the analytical Jacobian of the drift ($f'(x) = 1-3x^2$). The UKF utilized a standard symmetric sigma-point set ($\alpha=10^{-3}, \beta=2, \kappa=0$), while the Particle Filter employed $N=2,000$ particles with systematic resampling. A Regular Kalman Filter served as a baseline, statically linearized around the stable equilibrium ($x=1$), yielding a constant decay model. Table~\ref{tab:metastable-results1} summarizes performance over 40 independent trials run up to $t=100$.
The Lifted-KF achieves lower but comparable Root Mean Square Error (RMSE) than the other methods.

\begin{table}[h]
\centering
\small
\begin{tabular}{lcccc}
\toprule
Method & RMSE (mean $\pm$ std) & 95\% CI & Paired Diff vs. Lifted & Time per-trial (ms)\\
\midrule
Lifted-KF & 0.2132 $\pm$ 0.0051 & [0.2116, 0.2148] & &76.45\\  
EKF & 0.2239 $\pm$ 0.0054 & [0.2222, 0.2256] & -0.0107 $\pm$ 0.0006&3.38\\   
UKF & 0.2239 $\pm$ 0.0054 & [0.2223, 0.2256] & -0.0108 $\pm$ 0.0006&42.48\\ 
Particle Filter &  0.2240 $\pm$ 0.0053 & [0.2223, 0.2257] & -0.0108 $\pm$ 0.0006&683.82\\ 
Regular KF & 0.2246 $\pm$ 0.0054 & [0.2229, 0.2263] & -0.0114 $\pm$ 0.0004&1.09\\
\bottomrule
\end{tabular}
\caption{Comparative tracking performance in the Cubic Bistable process ($T=100$, 40 trials). RMSE statistics, a paired-difference analysis relative to the Lifted-KF, and average wall-clock time per trail. The Lifted-KF achieves accuracy comparable to or better than the benchmarks.}
\label{tab:metastable-results1}
\end{table}

\subsection{Radial (Bessel) Process}

The radial SDE is given in equation~\ref{eq:bessel_process}. We generate continuous-time trajectories on the bounded interval $[0,R]$ using a Euler--Maruyama step $dt$ (e.g.\ $dt=10^{-3}$), enforce reflecting boundaries at $0$ and $R$, and sample observations at interval $\Delta=0.1$. 
\[
  r_{t+dt} = r_t + \frac{(d-1)\sigma^2}{2r_t}\,dt + \sigma\sqrt{dt}\,\xi_t,
  \qquad 
  \xi_t \sim \mathcal{N}(0,1),
\]
followed by reflection if $r_{t+dt}<0$ or $r_{t+dt}>R$, and
\[
  y_k = r_{t_k} + \varepsilon_k,
  \qquad 
  \varepsilon_k \sim \mathcal{N}(0,\sigma_y^2).
\]
At each observation time $t_k = k\Delta$ we add Gaussian measurement noise of variance $\sigma_y^2$. Here also we take {$\sigma_y=0.25$}, the initial state $r_0$ is sampled from the stationary density $\rho(r)\propto r^{d-1}$ on $[0,5]$. 

Unlike the cubic bistable and Wright--Fisher examples, the Bessel process exhibited degraded filtering performance when additional exponential basis functions were included. An ablation study over (M=2,\ldots,5) therefore selected (M=2) for the tracking experiments reported in this section, and the corresponding lifted matrices ((A,B)) were recomputed using this reduced basis. We use one exponential basis function, $U(r) = (r, e^{\alpha r})$, parameterized by a single exponent~$\alpha$.

We discretize this model using the Van Loan method (to handle the stiffness near the origin) and obtain $(F,Q)$ \citep{vanloan1978computing}. We then apply the standard Kalman filter to the lifted states $\{U_k\}$. The filter is initialized by mapping $r_0$ to $U(r_0)$ and using a conservative diagonal covariance $\Sigma_{U,0}$ like in the previous case.

Here also we consider a regime with true noise $\sigma_x=2$ and assumed model noise $\sigma_m=1$. We compare the Lifted-KF to EKF, UKF, particle filter and a regular linear Kalman filter. 

For benchmarking, the EKF employed the analytical Jacobian of the drift $f'(r) = -(d-1)\sigma^2/2r^2$, which becomes singular at the origin. The UKF used the same sigma-point parameters as the bistable case ($\alpha=10^{-3}, \beta=2, \kappa=0$), but with a singularity protection floor ($r_{\min}=10^{-10}$) to prevent invalid evaluations during sigma-point propagation. The Particle Filter ($N=2,000$) implemented a reflecting boundary condition: particles predicted to cross $r=0$ or $r=R$ were reflected back into the domain to strictly enforce the state constraints. A Regular Kalman Filter provided a baseline by linearizing the drift at a mean radius $\bar{r} = dR/(d+1)$.

Table~\ref{tab:radial-results} summarizes performance over 40 independent trials run up to $t=100$. 
\begin{table}[H]
\centering
\small
\begin{tabular}{lcccc}
\toprule
Method & RMSE (mean $\pm$ std) & 95\% CI & Paired Diff vs. Lifted & Time per-trial (ms)\\
\midrule
Lifted-KF & 0.2380 $\pm$ 0.0055 & [0.2362, 0.2397]&  & 68.65\\
EKF & 0.2506 $\pm$ 0.0059 & [0.2488, 0.2525]& -0.0127 $\pm$ 0.0019& 5.19\\   
UKF & 0.2506 $\pm$ 0.0059 & [0.2488, 0.2525] &-0.0127 $\pm$ 0.0011& 47.09\\ 
Particle Filter &  0.2519 $\pm$ 0.0065 & [0.2498, 0.2539]& -0.0139 $\pm$ 0.0014& 857.96 \\ 
Regular KF & 0.2507 $\pm$ 0.0059 & [0.2488, 0.2525]& -0.0127 $\pm$ 0.0011 & 1.51\\ 
\bottomrule
\end{tabular}
\caption{Comparative tracking performance in the radial (Bessel) process ($T=100$, 40 trials). RMSE statistics, a paired-difference analysis relative to the Lifted-KF, and average wall-clock time per trail. The Lifted-KF achieves accuracy comparable to or better than the benchmarks.}
\label{tab:radial-results}
\end{table}

The baseline nonlinear filters (EKF/UKF) struggle on the radial Bessel process primarily because the true drift is extremely stiff for the parameter ranges used in our experiments. The Bessel drift
\[
f(r) = \frac{(d-1)\sigma^{2}}{2r}
\]
diverges like $1/r$ as $r\to 0$, a singularity known to destabilize standard discretization schemes \cite{alfonsi2005discretization}. For example, with $d=3$ and $\sigma\approx 5.5$, the linearization term in the EKF update,
\[
F_k = 1 + f'(r_k)\,\Delta = 1 - \frac{(d-1)\sigma^{2}}{2r_k^{2}}\,\Delta,
\]
often satisfies $|F_k|\gg 1$ for typical values of $r_k\in[0.3,1.0]$ and $\Delta=0.1$. Consequently, the EKF covariance update blows up rapidly, producing numerical instability and severe divergence. 

The lifted model is not derived by linearizing the nonlinear Bessel drift. Instead, we construct a linear surrogate in an augmented feature space. The penalty given in equation \ref{eq:constrained_optimization} enforces mean-square stability in lifted space, ensuring that the discretized dynamics $U_{k+1}=F U_k+w_k$ have eigenvalues inside the unit circle in the complex plane. Thus the lifted Kalman filter by construction does not experience the Jacobian blow-up that destabilizes the EKF/UKF.

However, the lifted model is still only an approximate representation of the nonlinear generator, and it captures the Bessel drift accurately only in an averaged sense with respect to the stationary density. As a consequence, the lifted filter is most accurate when the observations are informative. When the measurement noise $\sigma_y$ is small, the observation strongly constrains the lifted state $U_k$, preventing the filter from drifting into regions where the linear surrogate is less accurate. Thus the lifted filter's advantage is greatest when the measurement noise is small.

\subsection{Wright-Fisher Logistic Diffusion with Mutation}

As a third example we consider a bounded, multiplicative-noise diffusion on the
unit interval, given by the Wright-Fisher logistic-with-mutation SDE introduced
in Section~\ref{sec:Logistic}. We simulate continuous-time trajectories using an Euler--Maruyama step $dt=10^{-3}$, enforce numerical confinement to $(0,1)$ using reflective
clipping, and sample observations at interval $\Delta=0.1$:
\[
  X_{t+dt}
  = X_t
    + \kappa\!\bigl(\theta_1(1-X_t)-\theta_0 X_t\bigr)\,dt
    + \sqrt{2\kappa X_t(1-X_t)}\,\sqrt{dt}\,\xi_t,
  \qquad \xi_t\sim\mathcal{N}(0,1),
\]
followed by
\[
  y_k = X_{t_k} + \varepsilon_k,
  \qquad \varepsilon_k\sim\mathcal{N}(0,\sigma_y^2).
\]
As before we take $\sigma_y=0.25$. The model admits an explicit compact stationary
density, a Beta distribution
\[
  \rho(x)
  = \frac{1}{B(\theta_1,\theta_0)}
    x^{\theta_1-1}(1-x)^{\theta_0-1}, \qquad x\in(0,1),
\]
which we use to construct the weighted objective $J[U,A,B]$ for the lifted model.

In this example we use the exponential basis
$U(x)=(x,e^{\alpha_1 x},\dots,e^{\alpha_{M-1}x})$ with $M=4$.  As in the previous
sections, $(A,B)$ are learned by minimizing the stability-penalized objective,
and the resulting continuous-time dynamics are discretized exactly using the Van
Loan method to obtain $(F,Q)$ for the lifted linear system
$U_{k+1}=F U_k + w_k$.

We consider a regime with ``true'' simulation noise $\sigma_x=1$ and
assumed model noise $\sigma_m=1$ inside the EKF/UKF/linear-KF baselines.  The
Lifted-KF is initialized by mapping $X_0$ to $U(X_0)$ and using a conservative
diagonal prior covariance.  At each observation time we apply the standard
linear Kalman update in the $U$-coordinates and project back to $x_k=U_{0,k}$.

For benchmarking, the EKF propagated the mean using the already linear drift but calculated the process noise variance $Q_k \approx \sigma^2 x_k(1-x_k)\Delta$ based on the state-dependent diffusion at the current estimate. The UKF utilized standard parameters ($\alpha=10^{-3}, \beta=2, \kappa=0$). The Particle Filter ($N=2,000$) handled the domain constraints $[0,1]$ by simply clipping the particles after the Euler-Maruyama step, ensuring no probability mass leaked outside the valid interval. A Regular Kalman Filter served as a baseline, linearized around the deterministic equilibrium $x^* = \theta_1/(\theta_0+\theta_1)$.

Table~\ref{tab:wf-results} summarizes performance over 40 independent trials run up to $t=100$.

\begin{table}[h]
\centering
\small
\begin{tabular}{lcccc}
\toprule
Method & RMSE (mean $\pm$ std) & 95\% CI & Paired Diff vs. Lifted & Time per-trial (ms)\\
\midrule
Lifted-KF & 0.1550 $\pm$ 0.0036 & [0.1538, 0.1561] & & 76.27\\ 
EKF       & 0.1653 $\pm$ 0.0030 & [0.1643, 0.1662] & -0.0103 $\pm$ 0.0008& 29.58\\ 
UKF       & 0.1648 $\pm$ 0.0032 & [0.1638, 0.1658] & -0.0098 $\pm$ 0.0007& 72.40\\ 
Particle Filter &  0.1485 $\pm$ 0.0027 & [0.1477, 0.1493] & +0.0064 $\pm$ 0.0008 & 739.58\\ 
Regular KF & 0.1647 $\pm$ 0.0032 & [0.1637, 0.1658] & -0.0098 $\pm$ 0.0007& 1.34\\
\bottomrule
\end{tabular}
\caption{Comparative tracking performance in the Wright-Fisher process ($T=100$, 40 trials). RMSE statistics, a paired-difference analysis relative to the Lifted-KF, and average wall-clock time per trail.}
\label{tab:wf-results}
\end{table}

\begin{figure}[H]
    \centering
    \includegraphics[scale=0.5]{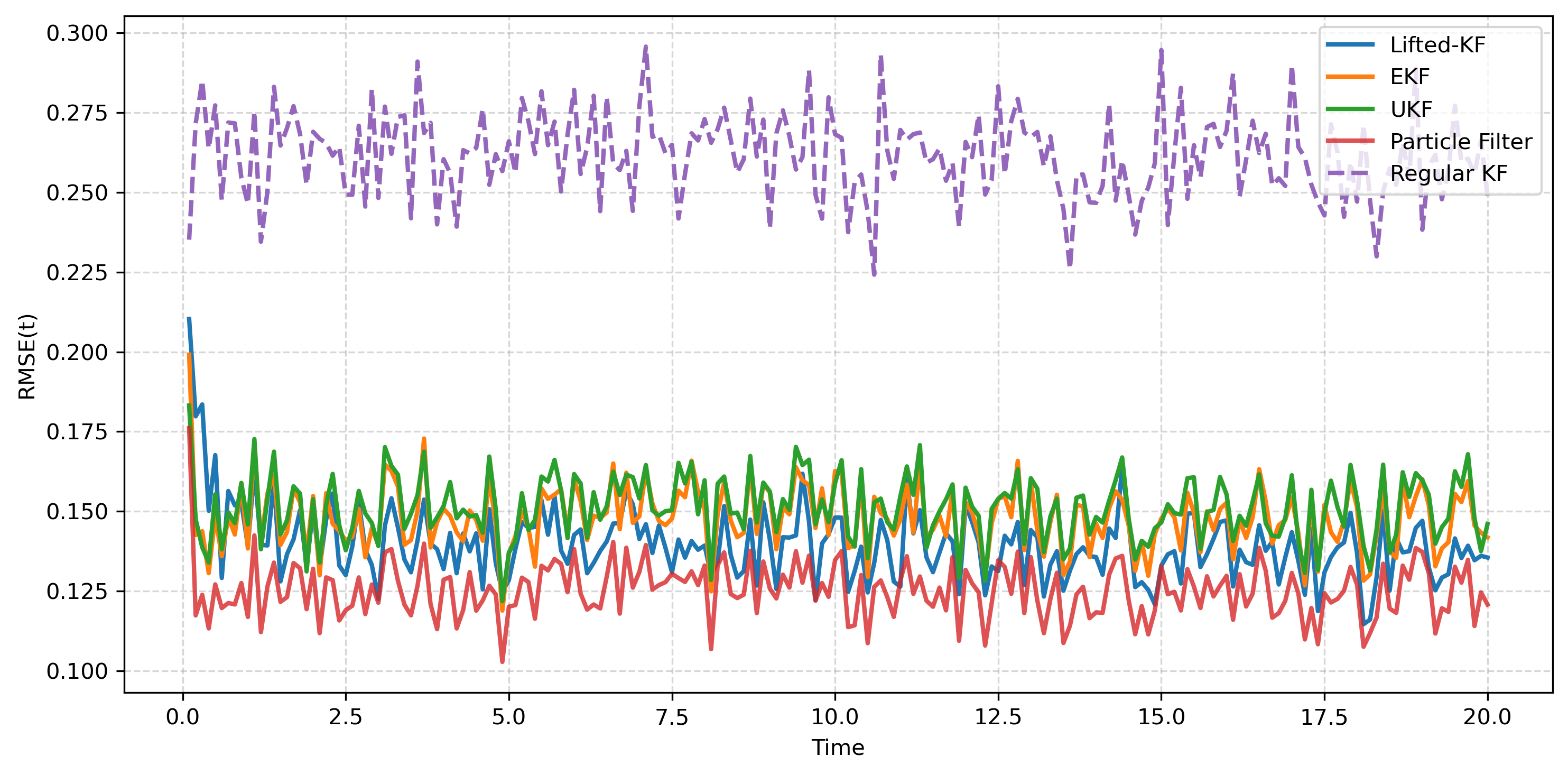}\\
    \caption{RMSE Time-Evolution for the Wright-Fisher case. Average tracking error over time ($T=20$) across 100 trials. The Lifted-KF (blue) maintains stable performance comparable to the computationally expensive Particle Filter (Red). The nonlinear baselines (EKF/UKF) exhibit higher error fluctuations and the regular KF exhibits the most error.}
    \label{fig:KF_compare_logistic}
\end{figure}

Next, we observe the RMSE across time of the various filtering methods for multiple trajectories in figure \ref{fig:KF_compare_logistic}. There we see under performance by EKF and UKF. These nonlinear filters face the structural challenge from the multiplicative diffusion
$\sqrt{2\kappa\,x(1-x)}$ vanishing at both boundaries, producing highly
state-dependent process noise that leads to severe underestimation or
overestimation of the predicted covariance if the linearization in the filters is even mildly misaligned.  

\section{Discussion}

We have presented in this work a dimension-lifting framework for the effective tracking of nonlinear noisy dynamics, typified by the singular drift fields of Bessel and Wright-Fisher processes. The framework relies on constructing a linear surrogate in a high-dimensional feature space, optimized via a variational principle. The major enabling factor for these results is the weighting of the optimization objective by the stationary density $\rho(x)$, which ensures that the learned dynamics prioritize fidelity in regions of high probability mass. Even with the suboptimal choice of exponential basis functions employed here, we obtain good agreement between the original and lifted systems, and the resulting tracking performance is decent. In the least accurate test case, the approximation fidelity of the cubic surrogate is sufficient for tracking at the observation density used here, but would be expected to degrade for sparser observations where prediction errors accumulate between updates.

A key finding of our study is the method's superior stability in systems with stiff or singular dynamics. In the Radial Bessel process, the local Jacobian diverges at the origin, causing standard approximate filters like the EKF and UKF to suffer from covariance inflation. Similarly, in the Wright-Fisher process, the vanishing diffusion at the boundaries leads to covariance collapse in these standard frameworks. Our method circumvents these structural pitfalls because the optimization objective $\mathcal{J}$ effectively filters out the singular behavior. Since $\rho(x)$ naturally vanishes or decays at these boundaries, the optimization algorithm automatically prioritizes the regions of state space where the system actually resides. This signifies that the linear surrogate captures the effective drift and diffusion averaged over the attractor, yielding a stable predictor even when the underlying generator is singular.

While the approximations we have made to obtain a tractable linear surrogate result in slightly higher RMSE compared to the Particle Filter in complex regimes (Table \ref{tab:wf-results}), the theoretical structure of the Lifted-KF offers substantial advantages. Tables 2–4 show that the Lifted-KF is not the fastest method overall, with the EKF, UKF, and regular KF typically requiring less computation per trial. However, the Lifted-KF consistently maintains stable tracking performance in regimes with singular or strongly nonlinear dynamics. Additionally, relative to the Particle Filter, the Lifted-KF substantially reduces computational cost by replacing the propagation and resampling of N=2000 particles with matrix-vector operations in the lifted state space. The tradeoff is that the Lifted-KF's offline optimization (BFGS) must be completed before deployment, which is feasible for systems where the underlying dynamics do not change abruptly.

While computationally efficient in one dimension, the scalability of the current implementation warrants further discussion. All three test cases considered here are one-dimensional. 
Extending the variational lifting framework to higher-dimensional state spaces introduces two key challenges. 
First, the number of basis functions grows combinatorially with dimension: for $M$ basis functions in $d$ dimensions, a natural product basis gives $M^d$ terms, making the variational optimization intractable beyond $d \sim 3$ without additional sparsity or independence assumptions. 
Second, the BFGS optimization of $(2M^2 + M - 1)$ parameters becomes increasingly expensive as $M$ grows. 
For $d$-dimensional systems, future work will investigate basis selection guided by other analytical approaches such as Koopman operator theory.

A limitation of the current lifting strategy is that the optimization may converge to exponent values that are nearly identical or close to zero. In such cases the resulting basis functions become nearly linearly dependent, producing an ill-conditioned lifted representation and potentially degrading filtering performance. This effect was observed most clearly in the Bessel example, where increasing the lift dimension improved the objective value but worsened the conditioning of the resulting state-transition matrix and ultimately increased tracking error. Developing basis-selection procedures that explicitly enforce orthogonality or rank preservation remains an important direction for future work.

Our primary purpose here was to demonstrate the feasibility of the lifting approach using simple exponential basis functions. In principle, a more systematic choice for future applications would be the family of orthogonal functions associated with the stationary measure $\rho(x)$, such as Jacobi polynomials for the Wright-Fisher process. Although these functions will not be exact solutions to the Euler-Lagrange equation, they guarantee that the variational approximation error vanishes asymptotically as the dimension of the lifted space goes to infinity. We leave the rigorous analysis of such orthogonal basis expansions and their convergence properties, as well as systematic basis-selection procedures in general for future work.

\section{Appendix: General Formulation} \label{appendix:GeneralFormulation}
Now we come back the general problem given in equation \ref{eq:J-continuous} and devise a general plan for a solution. When A and B are fixed, we get the Euler Lagrange equation of \(\mathcal{J}\) giving solutions of \textbf{U}. 
The first variation of \(\mathcal{J}\) is
\[
\begin{aligned}
\delta \mathcal{J}
&= 2\int_{\Omega} 
\big[
R^\top \big( \mathcal{L}(\delta U) - A\,\delta U  \big)
+ S^\top \big( G\,\delta U' - B\,\delta U \big)
\big]\, \rho(x)\,dx.
\end{aligned}
\]
Integrating by parts and moving the differential operator $\mathcal{L}$ onto $R$ gives
\[
\begin{aligned}
\delta \mathcal{J}
&= 2\int_{\Omega}
\delta U^\top
\Big[
\mathcal{L}^*(\rho R)
- A^\top(\rho R)
- \frac{d}{dx}\big(G\,\rho\,S\big)
+ B^\top(\rho S)
\Big]\,dx + \text{(boundary terms)}.
\end{aligned}
\]
where \(\mathcal{L}^*\) is the adjoint operator of \(\mathcal{L}\).
The requirement that the first variation vanishes (\(\delta \mathcal{J} = 0\)) provides the necessary condition for optimality, yielding the Euler-Lagrange system::
\begin{equation}\label{eq:EL-U}
\mathcal{L}^*\big(\rho\,R(x;U,A)\big)
- A^\top\big(\rho\,R(x;U,A)\big)
- \frac{d}{dx}\!\big(G(x)\,\rho(x)\,S(x;U,B)\big)
+ B^\top\big(\rho\,S(x;U,B)\big)
= 0
\end{equation}
Fully expanded we get:
\begin{equation}\label{eq:EL-U2}
\begin{aligned}
&-\frac{d}{dx}\Big( f^2(x)\rho(x) U' \Big)
    -\frac{d}{dx}\Big( \tfrac12 f(x)\rho(x) G^2 U'' \Big)
    +\frac{d}{dx}\Big( f(x)\rho(x) A U \Big)+ \tfrac12\frac{d^2}{dx^2}\Big( G^2 \rho(x) f(x) U' \Big)\\[4pt]
  &\quad + \tfrac14\frac{d^2}{dx^2}\Big( G^4 \rho(x) U'' \Big)
     - \tfrac12\frac{d^2}{dx^2}\Big( G^2 \rho(x) A U \Big) - A^\top\big(\rho(x) f(x) U'\big) - \tfrac12 A^\top\big(\rho(x) G^2 U''\big) \\[4pt]
  &\quad + A^\top\big(\rho(x) A U\big) - \frac{d}{dx}\big(G^2 \rho(x) U'\big) + \frac{d}{dx}\big(G \rho(x) B U\big)
    + B^\top\big(\rho(x) G U'\big) - B^\top\big(\rho(x) B U\big)=0 
\end{aligned}
\end{equation}

In particular, for the first two processes we considered above and in some regimes for the third process, we plug in the respective $f$, $G$, and $\rho$ and, after some cancellation, we get a system of polynomial coefficient linear ODEs given $A$ and $B$. The solution for $U$ will typically not be in standard mathematical functions and will require a numerical approach. 

Since the Euler–Lagrange equations lack closed-form solutions, we propose an alternating ``block-coordinate" strategy analogous to Expectation-Maximization. This approach bootstraps $(A,B)$ via the fixed-basis method, then iteratively updates $U$ (via a numerical solution to the Euler–Lagrange system) and re-optimizes $(A,B)$ (via BFGS). This generalizable framework is left for future work.
\\

\noindent\textbf{Declaration of generative AI and AI-assisted technologies in the manuscript preparation process.} During the preparation of this work the author used Claude in order to improve prose and organize code. After using this tool/service, the author reviewed and edited the content as needed and takes full responsibility for the content of the published article.

\bibliography{ref}
\bibliographystyle{unsrtnat}
\end{document}